\documentclass[conference]{IEEEtran}
\IEEEoverridecommandlockouts
\usepackage{cite}
\usepackage{amsmath,amssymb,amsfonts}
\usepackage{algorithmic}
\usepackage{graphicx}
\usepackage{textcomp}
\usepackage{xcolor}
\usepackage{hyperref}
\usepackage[nolist]{acronym}
\usepackage{paralist}
\def\BibTeX{{\rm B\kern-.05em{\sc i\kern-.025em b}\kern-.08em
    T\kern-.1667em\lower.7ex\hbox{E}\kern-.125emX}}

\begin{document}

\makeatletter
\def\ps@IEEEtitlepagestyle{%
\def\@oddfoot{\parbox{\textwidth}{\centering\normalsize{979-8-3503-9790-1/23/\$31.00 \copyright 2023 IEEE}\vspace{2em}}
}%
}

\makeatletter
\def\ps@IEEEtitlepagestyle{%
\def\@oddfoot{\parbox{\textwidth}{\footnotesize
Author's version of a paper accepted for publication in Proceedings of the 2023 International Conference on Smart Energy Systems and Technologies (SEST). 
\\
\textcopyright{} 2023 IEEE. 
Personal use of this material is permitted.  
Permission from IEEE must be obtained for all other uses, in any current or future media, including reprinting/republishing this material for advertising or promotional purposes, creating new collective works, for resale or redistribution to servers or lists, or reuse of any copyrighted component of this work in other works.\vspace{1.2em}}
}%
}
\makeatother

\acrodef{DSO}{Distribution System operator}
\acrodef{SCADA}{Supervisory Control and Data Acquisition}
\acrodef{DER}{Distributed Energy Resources}
\acrodef{ICT}{Information and Communication Technologies}
\acrodef{CIA}{Confidentiality, Integrity and Availability}
\acrodef{ICS}{Industrial Control System}
\acrodef{IT}{Information Technology}
\acrodef{OT}{Operational Technology}
\acrodef{MulVAL}{Multi-host, Multi-stage Vulnerability Analysis}
\acrodef{NVD}{National Vulnerability Database}
\acrodef{OVAL}{Open Vulnerability and Assessment Language}
\acrodef{TTC}{Time-to-Compromise}
\acrodef{IDS}{Intrusion Detection System}
\acrodef{CVE}{Common Vulnerabilities and Exposures}
\acrodef{CVSS}{Common Vulnerability Scoring System}
\acrodef{Ac}{Access Complexity}
\acrodef{Au}{Authentication}
\acrodef{Ex}{Exploitability}
\acrodef{DT}{Decision Tree}
\acrodef{GAN}{Generative Adversarial Network}
\acrodef{RF}{Random Forest}
\acrodef{SVM}{Support Vector Machine}
\acrodef{MCC}{Matthews correlation coefficient}
\acrodef{AUC}{Area Under Curve}
\acrodef{ROC}{Receiver Operating Characteristic}
\acrodef{TP}{True Positive}
\acrodef{TN}{True Negative}
\acrodef{FP}{False Positive}
\acrodef{FN}{False Negative}
\acrodef{CNB}{Complement Naïve Bayes}
\acrodef{XGB}{Extreme Gradient Boosting}
\acrodef{MQTT}{Message Queuing Telemetry Transport}
\acrodef{TTC}{Time-to-Compromise}
\acrodef{ML}{Machine Learning}
\acrodef{HMI}{Human-Machine-Interface}
\acrodef{CI}{Confidence Interval}

\bstctlcite{IEEEexample:BSTcontrol}

\title{Investigation of Multi-stage Attack and Defense Simulation for Data Synthesis}

\author{
\IEEEauthorblockN{%
Ömer Sen\IEEEauthorrefmark{1}\IEEEauthorrefmark{2},
Bozhidar Ivanov\IEEEauthorrefmark{1},
Martin Henze \IEEEauthorrefmark{3}\IEEEauthorrefmark{4},
Andreas Ulbig \IEEEauthorrefmark{1}\IEEEauthorrefmark{2},
}

\IEEEauthorblockA{%
\IEEEauthorrefmark{1}\textit{IAEW, RWTH Aachen University,} Aachen, Germany \&
\IEEEauthorrefmark{2}\textit{DE, Fraunhofer FIT} Aachen, Germany\\
Email: \{oemer.sen, andreas.ulbig\}@fit.fraunhofer.de, \{o.sen, a.ulbig\}@iaew.rwth-aachen.de, bozhidar.ivanov@rwth-aachen.de}
\IEEEauthorblockA{%
\IEEEauthorrefmark{3}\textit{SPICe, RWTH Aachen University,} Aachen, Germany \&
\IEEEauthorrefmark{4}\textit{CAD, Fraunhofer FKIE,} Wachtberg, Germany\\
Email: henze@cs.rwth-aachen.de}
}

\maketitle

\begin{abstract}
The power grid is a critical infrastructure that plays a vital role in modern society.
Its availability is of utmost importance, as a loss can endanger human lives.
However, with the increasing digitalization of the power grid, it also becomes vulnerable to new cyberattacks that can compromise its availability.
To counter these threats, intrusion detection systems are developed and deployed to detect cyberattacks targeting the power grid.
Among intrusion detection systems, anomaly detection models based on machine learning have shown potential in detecting unknown attack vectors.
However, the scarcity of data for training these models remains a challenge due to confidentiality concerns.
To overcome this challenge, this study proposes a model for generating synthetic data of multi-stage cyber attacks in the power grid, using attack trees to model the attacker's sequence of steps and a game-theoretic approach to incorporate the defender's actions.
This model aims to create diverse attack data on which machine learning algorithms can be trained.
\end{abstract}


\section{Introduction} \label{sec:introduction}
\vspace{-0.5em}
\subsection{Motivation \& Background}
The availability of power grids has become increasingly important in recent years due to the growing need for stable energy supply.
However, the increased usage of \ac{ICT} in distribution grids has introduced new possibilities and dangers, particularly in terms of cybersecurity attacks~\cite{b2}.
The 2015 cyberattack on Ukrainian regional Distribution System operator companies serves as an example of the potential disruption caused by such attacks~\cite{b1}.
With the increasing size and complexity of power grids, as well as the growing dependence on digitalization and renewable energy sources, safeguarding cybersecurity is essential to ensure the reliable functioning of power systems~\cite{lenanrt2023}.
\acp{IDS} are actively developed to enhance the protection of critical infrastructure by observing and scrutinizing network or system behavior to identify potential cyber threats~\cite{b3}.
Detecting and preventing planned attacks on the power system require the implementation of effective measures, and \ac{ML} algorithms are being developed for this purpose.
However, the lack of data can limit the predictability power of these algorithms, necessitating the generation of synthetic attack data that captures the characteristics of real attacks.
\vspace{-0.5em}
\subsection{Relevant Literature}\label{subsec:related_work}
Several approaches have been explored for generating synthetic cyberattack data for smart grids, including the creation of lab environments such as the CICIDS2017 dataset~\cite{sharafaldin2018toward}, the development of the ID2T framework for reproducible datasets~\cite{cordero2021generating}, modification of existing datasets like UNSW-NB15 using denoising autoencoders and Wasserstein \acp{GAN}~\cite{pandey2021gan}, and the generation of artificial attack data for smart grid security using frameworks like Melody~\cite{b4}.
Furthermore, studies such as~\cite{dutta2023deep} demonstrate the potential of data-driven deep reinforcement learning for proactive cyber defense, while~\cite{agnew2022implementation} introduces a cross-layered framework for securing the power grid.
Generating comprehensive and realistic datasets for effective ML model training is challenging due to various factors such as infrastructure implementation, scenario development, and ensuring data integrity and privacy.
Publicly available datasets may be limited and sharing data publicly can increase the risk of cyberattacks, highlighting the need for more comprehensive datasets.
This work utilizes game theory~\cite{b13} to model the dynamics between an attacker and defender in a power grid intrusion scenario, employing key terms such as ''starting capital'', ''funds'', ''betweenness centrality'', and ''path optimization'' to analyze their capabilities and strategic decision-making.
\vspace{-0.5em}
\subsection{Contributions \& Organization}
This paper aims to develop a model for generating synthetic data of cyberattacks for \ac{ML}-based \ac{IDS} in power grids, considering the dynamics between attackers and defenders.
Our contributions are:
\begin{enumerate}
\item We develop a model for generating synthetic cyberattack data for \ac{ML}-based \ac{IDS}.
\item We propose a method that incorporates attack tree modeling and game theory mechanics to generate diverse attack data.
\item We evaluate different ML models on the generated data and analyze the impact of attacker-defender dynamics on detection quality and attack complexity.
\end{enumerate}
The structure of this paper is as follows: Section~\ref{sec:framework} delves into the methodology used to generate synthetic attack data, followed by the evaluation of the generated data and the \ac{ML} models used for \ac{IDS} in Section~\ref{sec:result}.
The results are discussed, and conclusions are drawn in Section~\ref{sec:conclusion}.
\vspace{-0.5em}
\section{Nomenclature} \label{sec:nomenclature}
The mathematical symbols used in the equations throughout the paper are presented in Table~\ref{tab:nomenclature}.
\vspace{-4mm}
\begin{table}[h]
\centering
\caption{Nomenclature}
\begin{tabular}{p{1.7cm}p{6.5cm}}
\hline
Symbol & Description \\
\hline
$W_{i,j}$ & Weight of the edge connecting nodes $i$ and $j$ \\
$t$ & Time needed for a particular step \\
$C_{j}^{attacker}$ & Outage costs for the component from the attacker's viewpoint \\
$P_{j}^{attacker}$ & Likelihood of successfully compromising a node \\
$Risk$ & Risk of grid operation disruption due to cyberattack \\
$Q_{i}$ & Learning rate for each node $i$ \\
$c_{CB}(v)$ & Current flow betweenness centrality \\
$\tau_{st}(v)$ & Throughput from node $s$ to node $t$ via node $v$ \\
$b_{st}(v)$ & Absolute value of the total amount of current that flows through $v$ \\
$r(\vec{e_{i, j}})$ & Resistance between nodes $i$ and $j$ \\
$c_i ^ {outage}$, $c_j ^ {outage}$ & Outage costs of nodes $i$ and $j$ \\
$TTC(s, W)$ & Time to Compromise \\
$t_1$, $t_2$ & Time taken for the first and second stages of an attack \\
$P_1$ & Probability of the first stage of an attack \\
$u$ & Unsuccessful rate of the second stage of an attack \\
$N$ & Total number of nodes \\
\hline
\end{tabular}
\label{tab:nomenclature}
\end{table}
\begin{figure}\
\centerline{\includegraphics[width=\linewidth]{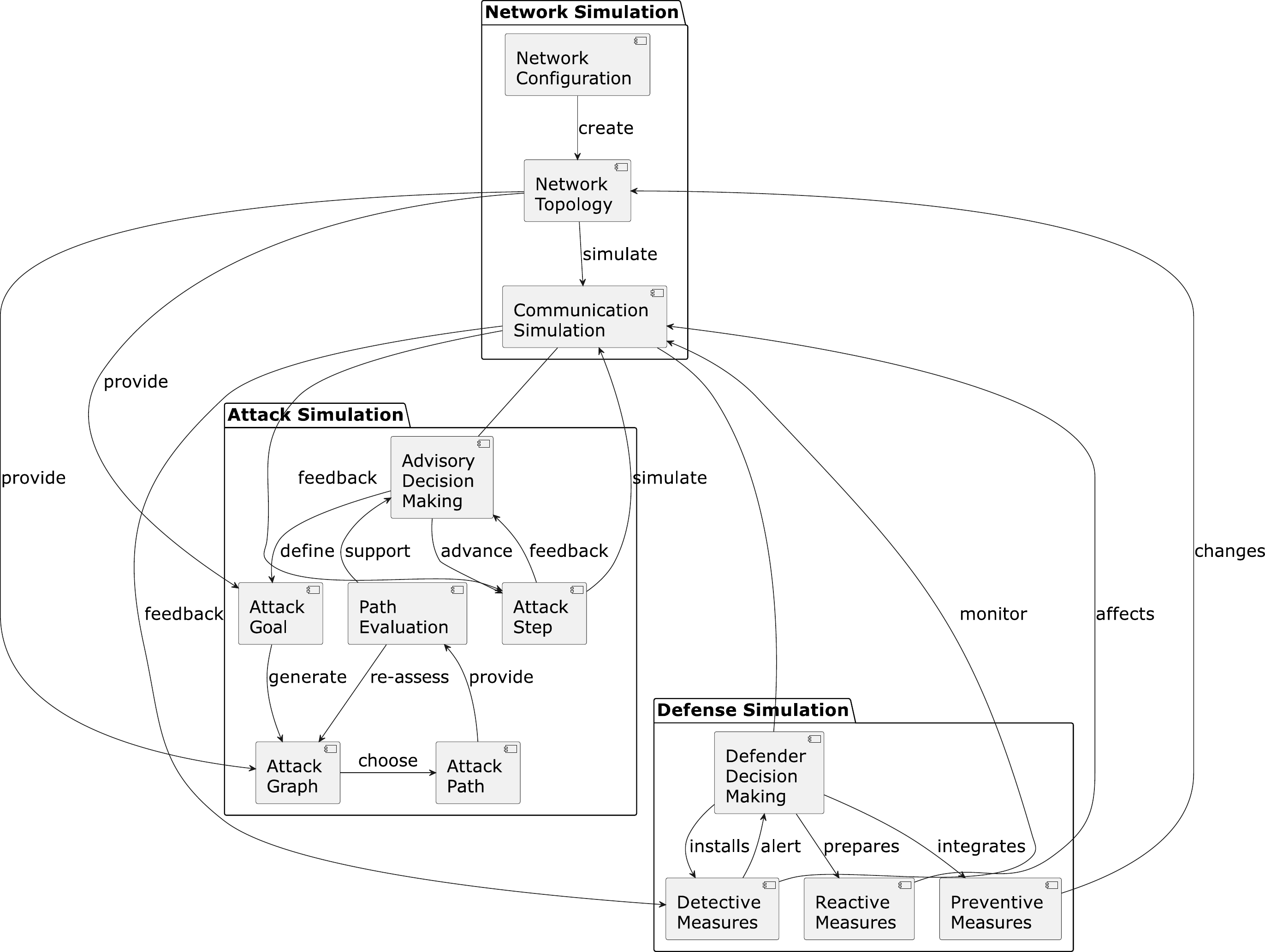}}
\caption{Structural overview of the presented approach, which includes a game-theoretic simulation of the dynamic interplay of cyberattacks and defenses, with a focus on continuous learning and strategic adaptation.}
\label{fig:framework_overview}
\vspace{-1.5em}
\end{figure}

\section{Multi-Staged Attack \& Defense Simulation} \label{sec:framework}
As our main contribution in this work, we present our method for generating synthetic cyberattack data using a game-theoretic approach between the attacker and defender.
\vspace{-0.5em}
\subsection{Overview} \label{subsec:framework_overview}
Our work is based on a game-theoretic approach for modeling the dynamics of cyberattacks and defenses (cf. Figure~\ref{fig:framework_overview}).
To understand how the dynamics between the attacker and defender affect data quality, we model a red vs. blue teaming approach in the simulation.
The attacker aims to cause disruption to the grid operation, while the defender aims to prevent it.
The attacker scans the system infrastructure and selects a target with high outage costs, while the defender takes preventative measures to lower the risk.
We utilize procedural attack graph generation, which adapts at each game turn considering the defender's counter actions.
The success rate is calculated based on prior knowledge, and the attacker chooses the shortest path using Dijkstra's Algorithm.
Both the defender and attacker have learning rates that influence their strategies, and the attack path evolves over time.
This learning effect is simulated by transferring knowledge from the previous simulation round to the next one.
Critical network nodes were equipped with \ac{IDS}, generating alerts for potential threats in a Python-based environment, where \ac{MulVAL} execution was facilitated via Docker containers.
The Python packages sklearn and xgboost were used to train \ac{ML} models, with optimal hyperparameters determined via Grid Search to minimize overfitting.
\vspace{-0.5em}
\subsection{Attack Graphs}
Modeling an attack on a complex system like the power grid requires considering multi-stage attacks instead of just single compromises.
Attack graph generation tools can model intricate situations involving simultaneous exploitation of multiple vulnerabilities, leading to a multi-stage cyberattack.
These tools also take into account the environment, severity, and impact of the exploited vulnerabilities.
Various attack graph generation tools were considered, with open-source tools providing more detailed attack paths but lacking user-friendliness due to poor visualization and obscurity.
To ensure detailed attack graphs and scalability, open-source tools like \ac{MulVAL}~\cite{b10} were chosen for their scalability and extensibility.
\ac{MulVAL} is a logical attack graph generation tool that uses Datalog as an input language~\cite{b10}.
It utilizes existing vulnerability databases such as the \ac{NVD}, allowing vulnerabilities to be specified by their ID.
Host and network configurations can be provided by an \ac{OVAL} scanner and a firewall management tool, respectively~\cite{b10}.
\vspace{-0.5em}
\subsection{Attacker and Defender}\label{subsec:attacker_defender}
\begin{figure}
\centerline{\includegraphics[width=\columnwidth]{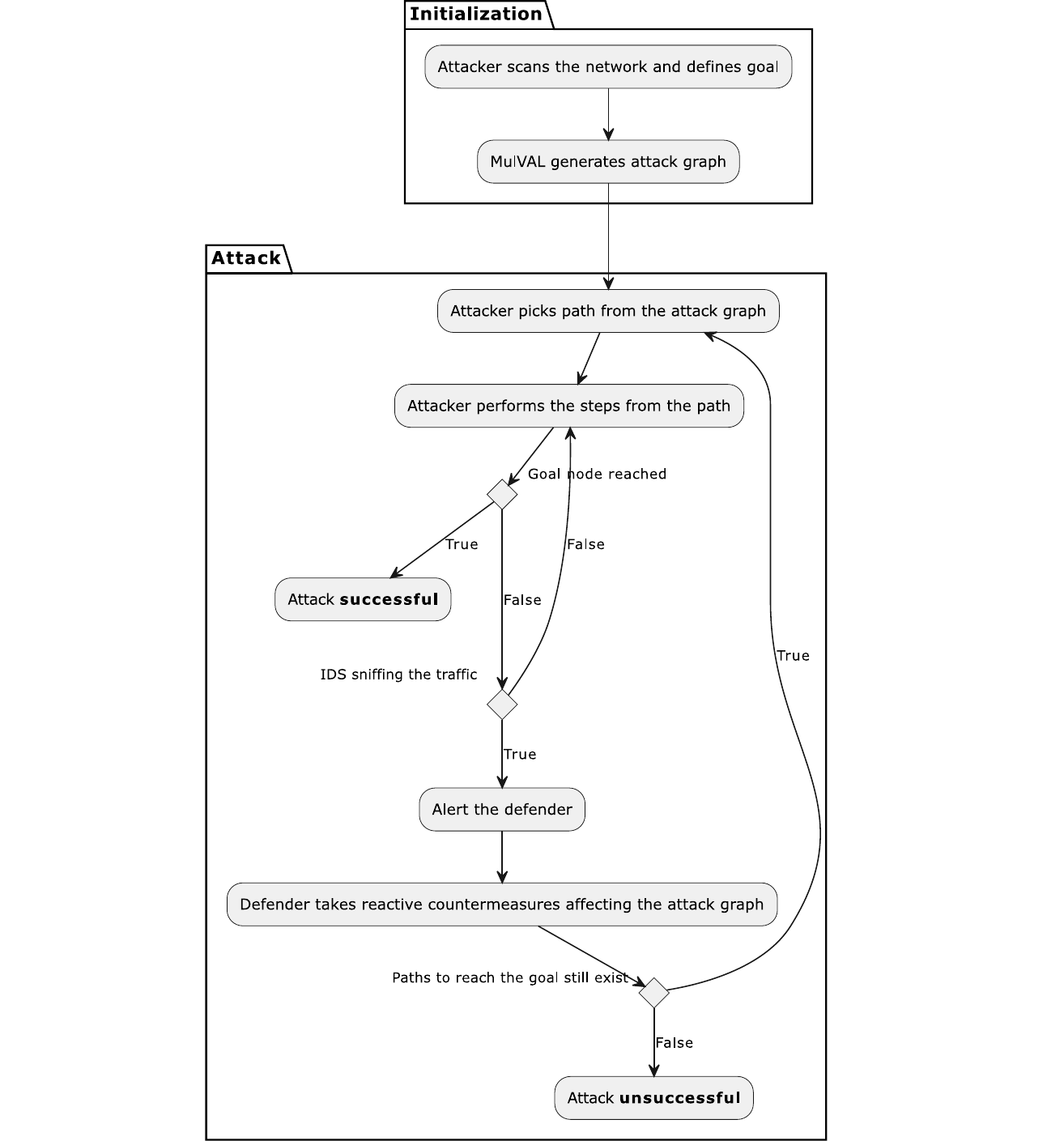}}
\caption{Overview of the attacker-defense dynamic, showcasing the attacker scanning the network and the defender employing preventive and reactive measures, including signature-based \ac{IDS} sensors for detection in the simulation.}
\label{fig:attack_defense}
\vspace{-1.5em}
\end{figure}

The attack and defense dynamics are modeled using a game-theoretic approach similar to~\cite{b13}.
The interaction between the attacker and defender is illustrated in Figure~\ref{fig:attack_defense}.
We model the actions of the attacker and defender using the MITRE ATT\&CK~\cite{strom2018mitre} and D3FEND~\cite{kaloroumakis2021toward} matrix.
Initially, the attacker performs a network scan to identify the node with the highest outage costs.
For instance, the \ac{SCADA} Server is highly susceptible to significant outage costs due to its expensive replacement and the potential for causing grid outages by compromising its security.
The defender takes preventative and reactive measures to reduce the risk (cf.~\cite{kaloroumakis2021toward}).
For example, the defender can use \ac{IDS} as sensors or employ access restriction measures.
\ac{IDS} sensors are placed between devices to detect attacks.
In the simulation, signature-based \ac{IDS} sensors are used to represent the defender's detection capability, classifying non-conforming data as attack-generated based on predefined rules.
These \ac{IDS} sensors are solely used to represent the defender's detection scale in the simulation and are not considered in the later evaluation, where ML-based anomaly detection scores are used instead.
The number of sensors is varied to observe its effect on the generated data.
The \ac{MulVAL} attack graph is created and transformed to represent the attacker's actions.
The attacker follows a path, compromising components until either they reach their goal or an \ac{IDS} detects their actions, triggering reactive countermeasures by the defender.
This process continues until the attacker succeeds or the attack graph becomes non-traversable, resulting in an unsuccessful attack.
Both the attacker and defender learn from previous experiences and take more optimal steps to improve the quality of attacks and overall strategy.
\vspace{-0.5em}
\subsection{Attacker}
\label{subsec: Attacker}
The attacker aims to cause disruption to the grid operation while minimizing the risk of being detected by an \ac{IDS}.
The disruption caused to the grid operation is calculated based on the outage costs caused by the attacker's actions.
To incrementally improve, the attacker increases their skill rate after each attack and updates their knowledge of the success rate of compromising a node.
The skill rate determines the success of an attack, considering failures and action attempts, and is probabilistic in nature.
Both factors influence the path chosen by the attacker, who uses Dijkstra's Algorithm to determine the optimal route~\cite{b11}.
The edge weights in the attack graph are computed using Equation~\ref{eq:edge_weight}, where $W_{i,j}$ represents the weight of the edge connecting nodes $i$ and $j$:
\begin{equation}
W_{i, j} = \frac{t_{j}^{attacker}}{C_{j}^{attacker} \cdot P_{j}^{attacker}}
\label{eq:edge_weight}
\vspace{-0.5em}
\end{equation}
The superscript ''attacker'' indicates that the components of the equation are evaluated from the attacker's viewpoint.
The variable $t$ represents the time needed for a particular step, determined by the \ac{TTC} metric outlined in Section~\ref{subsec: Risk evaluation}.
$P_{j}^{attacker}$ illustrates the likelihood of successfully compromising a node.
Initially, the attacker is unaware of the actual success rate and assumes it to be 1.
After each compromise attempt in both the current and previous attacks, a 1 or 0 is recorded to indicate success or failure, respectively.
The mean of these values is then used for the calculation.
If a random number between 0 and 1 is less than the actual success rate, the action can be executed by the attacker;
otherwise, they must retry, spending more time and increasing the defender's resources, which allows for the implementation of additional preventative measures, further reducing the success rate.
Alternatively, the attacker may reconsider the path if deterred by the updated success rate.
The outage costs for the component from the attacker's perspective are denoted by $C_{j}^{attacker}$.
\vspace{-0.5em}
\subsection{Defender}
\label{subsec: Defender}
The defender's goal is to minimize the risk of grid operation disruption due to cyberattacks.
They can implement preventative or reactive countermeasures, with preventative measures requiring time and funds to implement and operate, executed in subsequent simulation runs, while reactive measures can be executed instantaneously without capital, within the current run.
The starting capital and funds for the defender represent the initial and acquired financial resources used to implement defensive measures against cyber threats.
The defender enhances their knowledge by considering the attacker's most frequent pathways from previous attacks.
This influences risk calculation and the corresponding preventative countermeasures.
The calculation of the risk is defined in Equation~\ref{eq:risk_learn}, where the learning rate $Q_{i}$ is added for each node $i$ in the network topology graph, initially set to 1 and increasing with each attack if detected:
\begin{equation}
Risk = \sum_{i = 1}^N P_i \cdot C_i \cdot Q_{i}
\label{eq:risk_learn}
\vspace{-0.5em}
\end{equation}
\ac{IDS} sensor placement prioritizes nodes with higher outage costs.
Centrality-based algorithms, such as current flow betweenness centrality, are used to identify important graph elements~\cite{b15}.
Betweenness centrality measures a node's contribution to information exchange, assuming information flows along the shortest path.
Current flow betweenness centrality, which views the graph as an electrical network, removes this assumption:
\begin{equation}
c_{CB}(v) = \frac{1}{(n - 1) \cdot (n-2)} \sum_{s, t \in V} \tau_{st} (v)
\label{eq:current_flow_betw_centr}
\vspace{-0.5em}
\end{equation}
\begin{equation}
\tau_{st} (v) = \frac{1}{2}( - |b_{st}(v)| + \sum_{e \ni V} |x(\vec{e})|)
\label{eq:throughput}
\vspace{-0.5em}
\end{equation}
Equation~\ref{eq:current_flow_betw_centr} calculates the normalized sum of current flowing through node $v$ using $\tau_{st}(v)$ and a normalizing constant.
Equation~\ref{eq:throughput} defines the current flow using $b_{st}(v)$, and it ensures that the sum of all $b_{st}(v)$ is equal to 0 while considering the resistance of edges in the centrality measure.
Edge resistance between nodes $i$ and $j$ is set according to Equation~\ref{eq:resistance}, allowing easier current flow and higher centrality along edges connecting nodes with higher outage costs:
\begin{equation}
r(\vec{e_{i, j}}) = \frac{1}{\max(c_i ^ {outage}, c_j ^ {outage})}
\label{eq:resistance}
\vspace{-0.5em}
\end{equation}
The weighting of the edges is updated throughout the simulation runs, incorporating experiences from previous runs into the placement strategy.
Thus, the defender adjusts its sensor placement based on the previously observed effects of the attacker's actions.
\vspace{-0.5em}
\subsection{Steering Actions}
\label{subsec: Risk evaluation}
For the game-theoretic interaction between the attacker and defender, a target value is needed against which the attacker and defender can measure the yield of their actions.
The expected return is calculated based on risk assessment methodologies for cyberattacks on \ac{SCADA} systems, considering vulnerability, threat/attack, countermeasure, and impact~\cite{b18}.
Risk is a target value to be minimized from the defender's perspective, aiming to avoid or reduce potential damage, or maximized from the attacker's perspective, aiming to achieve the target.
The $\beta$-\ac{TTC} Metric for Practical Cyber Security Risk Estimation considers both vulnerabilities and attacker skills to estimate the time it takes to compromise a system~\cite{b19}.
Based on this, the \ac{TTC} and the success rate of an attack against each component $P_j$ are calculated using the model from~\cite{b19}.
The model requires vulnerability data from open-source datasets such as the \ac{CVE}, making it compatible with \ac{MulVAL} requirements.
\begin{equation}
TTC(s, W) = t_1 \cdot P_1 + t_2 \cdot (1 - P_1) \cdot (1 - u)
\label{eq:bttc}
\vspace{-0.5em}
\end{equation}
Impact is represented by the outage costs of each component $C_i$, calculated based on their importance to the grid operation.
The components' outage costs are determined using the Purdue model~\cite{b6} and the customer interruption cost of unplanned outages by peak power consumption for the industry sector from the model in~\cite{b22}.
A worst-case scenario of a 12-hour outage is considered.
The risk is calculated using Equation~\ref{eq:risk}:
\begin{equation}
Risk = \sum_{i = 1}^N P_i \cdot C_i
\label{eq:risk}
\vspace{-0.5em}
\end{equation}
Based on this dynamic interaction between the attacker and defender, we can simulate a logically bound chain of events within a cyber incident involving various actions.
This simulation approach is then used to generate synthetic data of cyber incidents, considering flexible scenarios between the attacker and defender.

\section{Results} \label{sec:result}
In this section, we present the results of the proposed training procedure for \ac{IDS} based on \ac{ML} using synthetic attack data.
\vspace{-0.5em}
\subsection{Investigation Procedure} \label{subsec:investigation_procedure}
The goal of this research is to understand the impact of complex patterns in attack data and investigate the effect of different levels of complexity on data quality.
To achieve this, we conducted an investigation using a multi-layer network architecture based on the Purdue model~\cite{b6}.
The architecture consists of 21 subnets representing a smart grid within an industrial control context.
Various countermeasures of the defender and capabilities of the attacker were integrated at different levels.
For example, we varied the coverage of \ac{IDS} sensors that sniff corporate network traffic, directly influencing attack propagation.
Using 5 to 15 sensors strikes an optimal balance between generating enough alarms from malicious activity and avoiding excessive defender intervention in our case studies.
Additionally, we manipulated parameters affecting the speed and spread of the attack, enabling the generation of complex datasets.
We evaluated data quality and its variation by training selected \ac{ML}-based anomaly detection models on log files generated during attacks in the Unified2 \ac{IDS} event format~\cite{au2016multi} (cf. Table~\ref{tab:gen_alerts}).
These are the input features for the \ac{ML} models, which aim to predict whether the alerts generated are from background processes or due to the attacker's actions based on the format of the alerts.
The best performing models were selected and compared using the evaluation method.
Furthermore, we analyzed the influences and expression of data quality with respect to parameterization and presented the findings accordingly.
The investigation was conducted on a PC equipped with a multicore CPU, at least 16GB RAM, a dedicated GPU with 8GB+ VRAM, and fast storage, such as SSD.
The investigation involved synthetic data generation, game-theoretic interactions, path optimization, and attack graph generation with $O(N^3)$ complexity.
\vspace{-1em}
\begin{table}[ht]
\centering
\caption{Generated Alerts from the tool in Unified2 format}
\label{tab:gen_alerts}
\begin{tabular}{llll}
\hline
label & size & label & size \\ \hline
sensor id & 4 bytes & source port/icmp type & 2 bytes \\
event id & 4 bytes & dest. port/icmp code & 2 bytes \\
event second & 4 bytes & protocol & 1 byte \\
event microsecond & 4 bytes & impact flag & 1 byte \\
signature id & 4 bytes & impact & 1 byte \\
generator id & 4 bytes & blocked & 1 byte \\
signature revision & 4 bytes & mpls label & 4 bytes \\
classification id & 4 bytes & vlan id & 2 bytes \\
priority id & 4 bytes & padding & 2 bytes \\
ip source & 16 bytes & application id & NA \\
ip destination & 16 bytes & sequence number & NA \\ \hline
\end{tabular}
\end{table}
\vspace{-0.5em}
\subsection{Complexity}
\begin{figure}[ht]
\centerline{\includegraphics[width=\columnwidth]{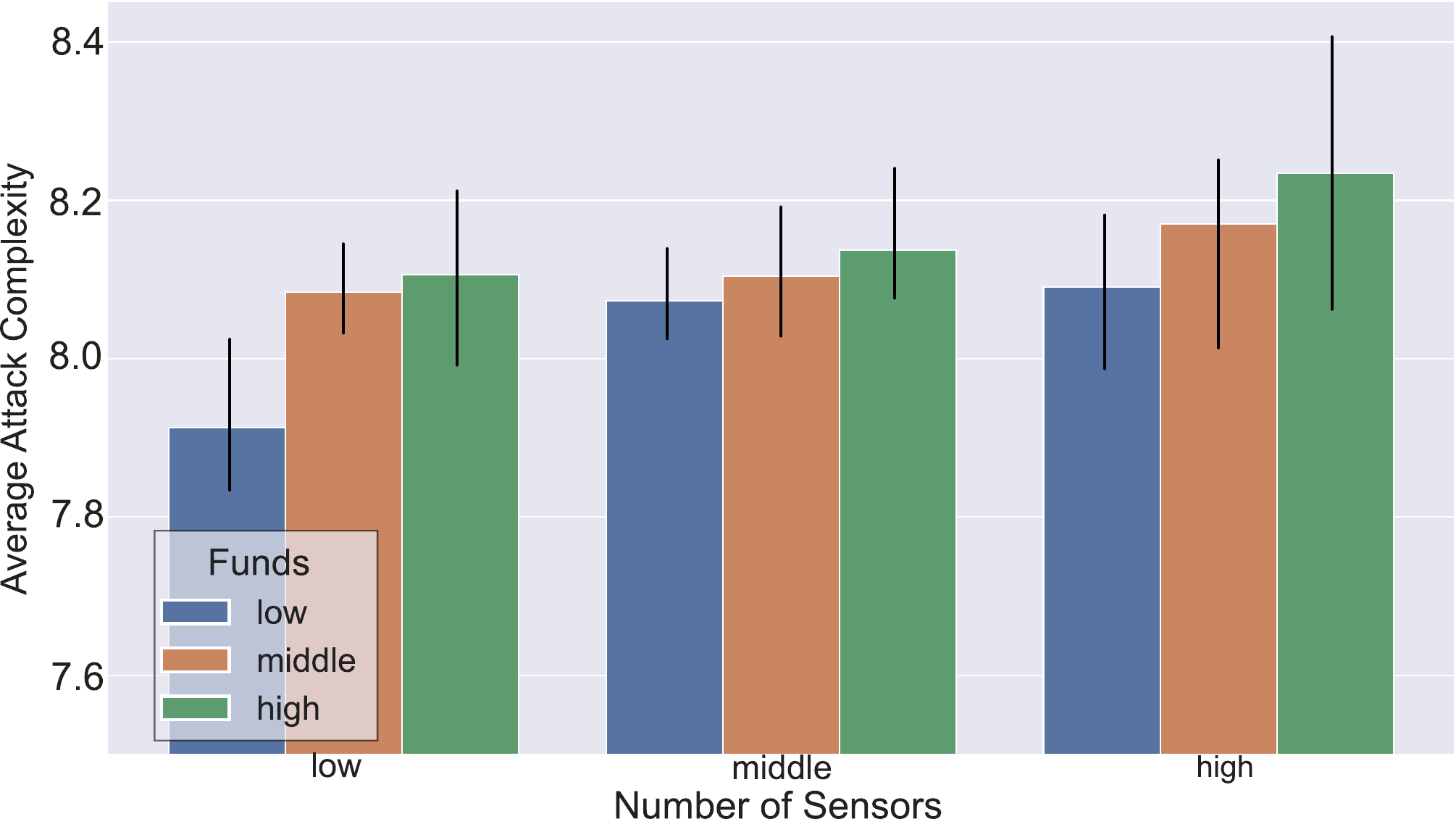}}
\caption{Attack Complexity of the generated attacks with different start configurations.
The y-axis represents the average complexity score over the simulation runs, while the error bar represents the 95\% \ac{CI}.
The x-axis represents the number of placed \ac{IDS} sensors, and the color labels of the bars represent the investment of the defender in preventive measures.}
\label{fig:attack_complexity}
\end{figure}

To study the diversity of actions, we measured the complexity of the attack on a scale from 0 (non-existent) to 10 (very complex attack) using the \ac{CVSS}~\cite{b25} and the length of the propagation path.
Different start configurations were used to test \ac{ML} models on 30 consecutive attacks (cf. Figure\ref{fig:attack_complexity}).
The attacker's skill level started at 0.5 and was incremented by 0.02 until reaching 1 after the 25th attack.
We considered three different scenarios for the starting capital and the funds gained per second.
The number of sensors was also varied to manipulate the difficulty of the attacks.
The complexity of the attacks was calculated based on the exploit complexity of the vulnerabilities used.
Figure~\ref{fig:attack_complexity} shows the average complexity score of the exploited vulnerabilities for all 30 performed attacks with a 95\% \ac{CI}.
Increasing the number of \ac{IDS} sensors and the amount of funds led to higher attack complexity.
Complexity also increased when the attacker was required to take alternative paths after reactive countermeasures were deployed or when more vulnerabilities were exploited.
\vspace{-0.5em}
\subsection{Performance}\label{subsec:performance}
\begin{figure}[ht]
\centerline{\includegraphics[width=\columnwidth]{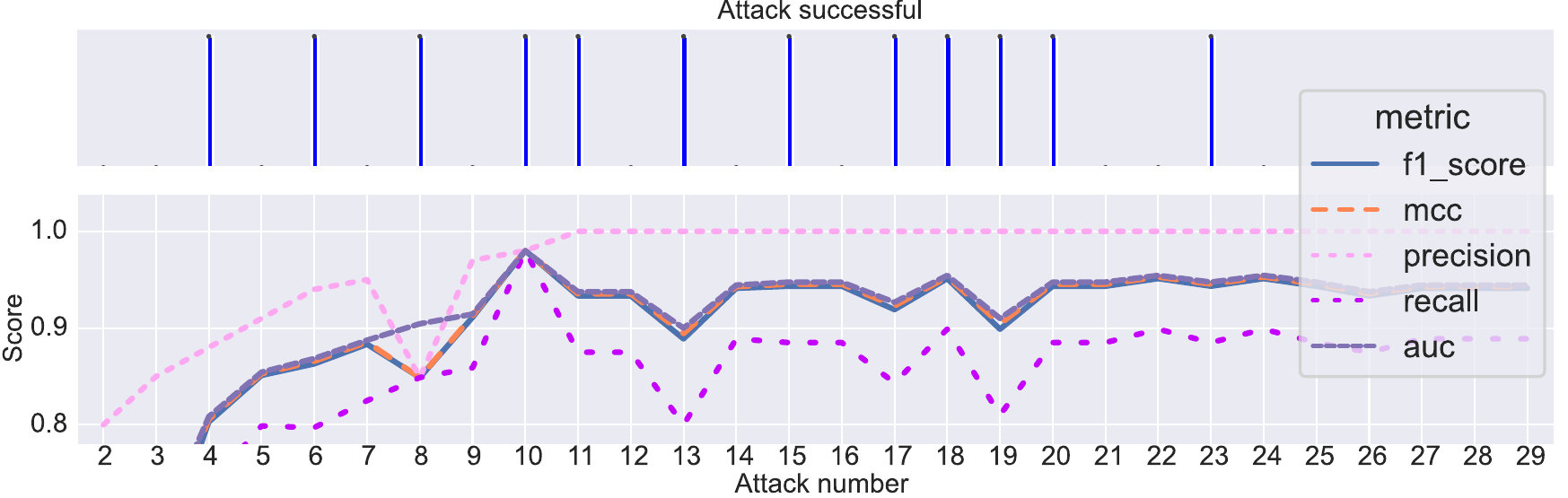}}
\caption{Evaluation of the \ac{RF} model.
The x-axis represents the simulation iteration, while the color labels of the lines represent different metrics.}
\label{fig:evaluation_rf}
\vspace{-1em}
\end{figure}

\begin{figure}[ht]
\centerline{\includegraphics[width=\columnwidth]{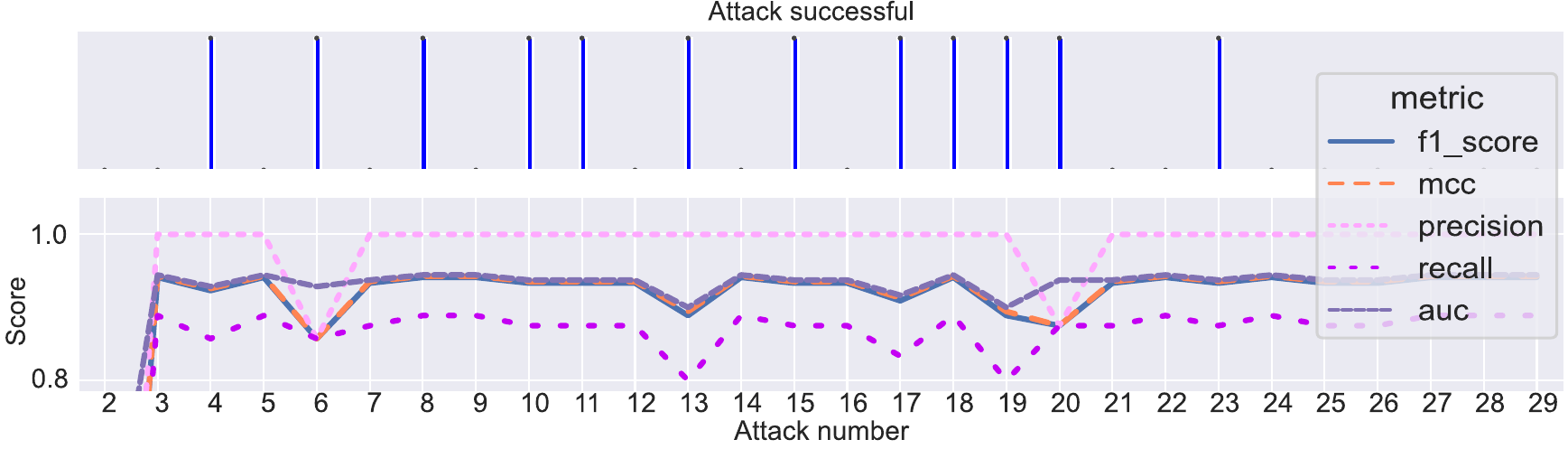}}
\caption{Evaluation of the \ac{XGB} model.
The x-axis represents the simulation iteration, while the color labels of the lines represent different metrics.}
\label{fig:evaluation_xgb}
\vspace{-1em}
\end{figure}

\begin{figure}[ht]
\centerline{\includegraphics[width=\columnwidth]{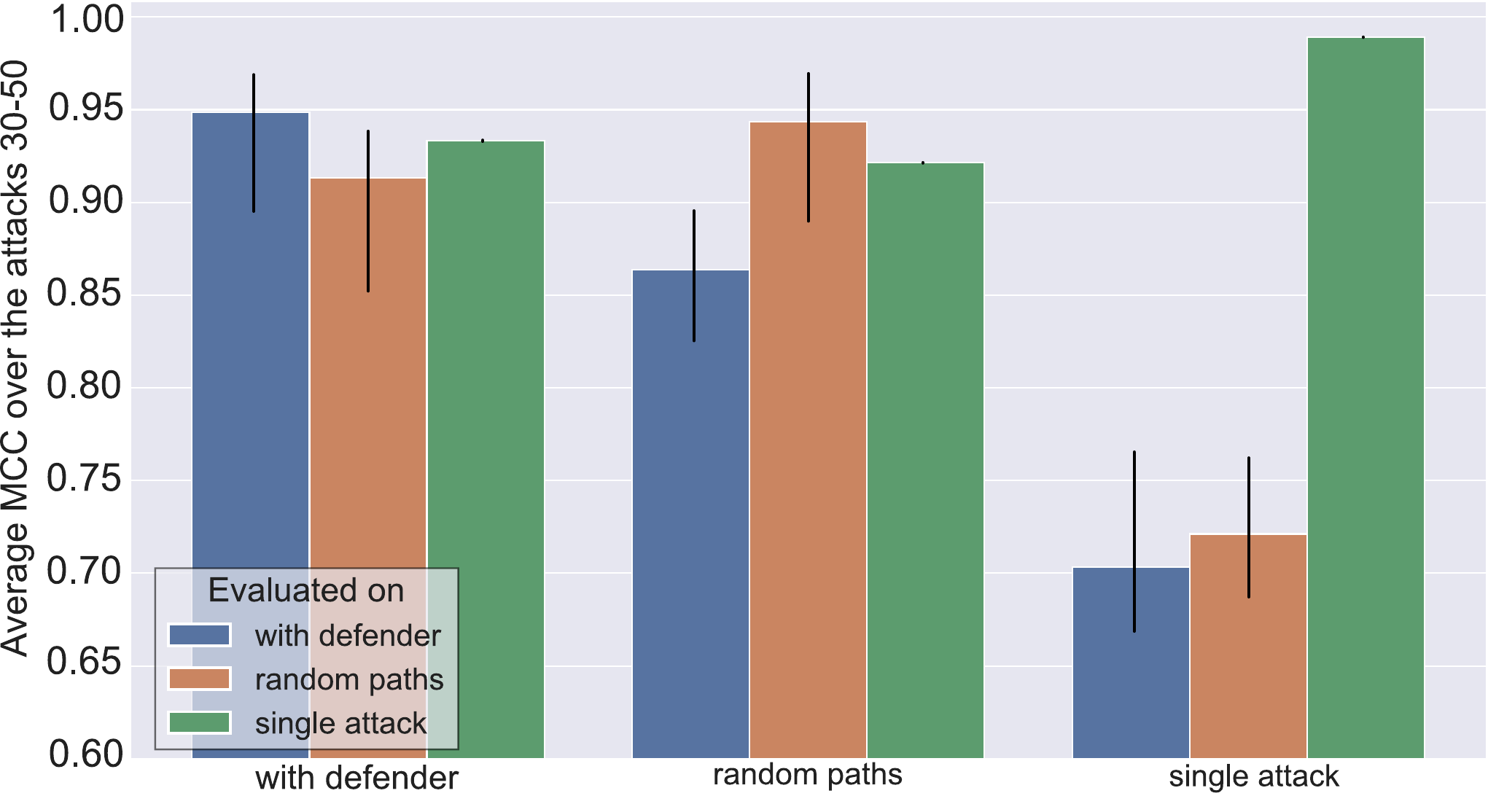}}
\caption{Evaluation of the different generation methods.
The y-axis represents the \ac{MCC} over simulation runs 30-50, while the error bar represents the 95\% \ac{CI}.
The x-axis represents the training data, and the color labels of the bars represent the test data.}
\label{fig:evaluation}
\vspace{-1.5em}
\end{figure}

To select an appropriate \ac{ML} algorithm for classification, we considered techniques presented in the \ac{IDS} review~\cite{b25} and a referenced paper~\cite{b26}.
From the supervised learning methods, \ac{RF}, \ac{DT}, \ac{SVM}, \ac{CNB}, and \ac{XGB} were chosen.
CNB was preferred over the Gaussian variant due to its better performance with imbalanced datasets~\cite{b27}.
K-Means was used as an unsupervised learning algorithm.
Based on previous research~\cite{b25,b26}, we chose the K-Means algorithm for time series data and used Hyper Parameter Grid Search to select the optimal value of k, considering performance and minimal overfitting observed in our experiments.
The models were iteratively evaluated against the data generated by the simulation runs, with cumulative data from previous runs used for training and the current iteration for testing.
The model parameters were optimized to balance overfitting and predictability.
The results were presented based on a scenario with a medium number of sensors and funds.
Although other scenarios were tested, the performance ranking of the models only showed minor differences.
Table~\ref{tab:models_scores} reports the performance of each \ac{ML} technique for the last simulation iteration (30th), including various classification metrics such as accuracy, $F_{1}$-score, \ac{AUC}, and \ac{MCC}, which are useful for imbalanced datasets.
Notably, \ac{SVM} exhibited slow fitting time with large amounts of data, as observed in~\cite{b28}, and was not useful for the evaluation method described in Subsection~\ref{subsec:investigation_procedure}.
The K-Means model was found to be the worst performing, while the \ac{XGB} model performed the best based on training time and performance metrics.
The performance of the \ac{RF} (cf. Figure~\ref{fig:evaluation_rf}) and \ac{XGB} (cf. Figure~\ref{fig:evaluation_xgb}) models over simulation runs 1-29 are depicted in graphs showing the number of attacks and their success rates.
Both models initially struggled with recognizing specific attacks but eventually gained an understanding of the attack structure.
It is worth noting that there were no successful attacks after the 23rd one, and the attacker's skill level reached a plateau of 1 after the 25th attack.
We further evaluated the defender implementation through three different methods of generating attack data (cf. Figure~\ref{fig:evaluation}).
The simulation runs were extended to 50 iterations to ensure consistency in the results.
The first method involved interaction with the defender, similar to the previous section.
In the second method, called ''single attack'', the attacker chose random paths without the defender taking countermeasures.
The third method involved the attacker taking the most optimal path without any defender implementation.
\ac{XGB} models were trained on attack data from attacks 1-29 and tested on attacks 30-50, and each model was tested on data from the other generation methods.
The results of the comparison between the three scenarios, namely ''random'', ''single attack'', and ''with defender'' revealed that the training data generated with the defender resulted in significantly improved detection quality.
This can be attributed to the increased diversity and the presence of plausible attack patterns that were not as pronounced in the random or single attack scenarios, as the defender's countermeasures forced the attacker to alter their strategy from the optimal short path.
\vspace{-1em}
\begin{table}[ht]
\centering
\caption{Scores of the different \ac{ML} models using various metrics}
\begin{tabular}{|p{1.3cm}|p{0.75cm}|p{0.75cm}|p{0.75cm}|p{0.75cm}|p{0.75cm}|p{0.75cm}|}
\hline
Metric & \ac{RF} & \ac{DT} & \ac{SVM} & \ac{CNB} & K-Means & \ac{XGB} \\ \hline
Accuracy & 0.9375 & 0.8182 & 0.9382 & 0.6697 & 0.5003 & 1 \\ \hline
Recall & 0.8889 & 0.8889 & 0.89 & 0.6810 & 0.0003 & 0.8889 \\ \hline
Precision & 0.9999 & 0.7273 & 0.9889 & 0.7 & 0.5555 & 1 \\ \hline
$F_{1}$-Score & 0.9411 & 0.8000 & 0.9369 & 0.6904 & 0.0006 & 0.9412 \\ \hline
\ac{AUC} & 0.9444 & 0.8391 & 0.9251 & 0.6841 & 0.5279 & 0.9444 \\ \hline
\ac{MCC} & 0.9333 & 0.6471 & 0.8721 & 0.3367 & 0.0005 & 0.9428 \\ \hline
\end{tabular}
\label{tab:models_scores}
\end{table}
\vspace{-0.5em}
\subsection{Discussion}
During the investigation, we observed that increasing the number of sensors and available funds for defense led to more complex attacks.
This occurred because the attacker was forced to take different, more difficult paths with complex vulnerabilities.
This could happen either because the original path was blocked by sensors and reactive countermeasures from the defender or because the attacker became discouraged by the low success rate of their actions due to increased preventative countermeasures requiring more funds.
The evaluation also revealed that the predictive ability of \ac{ML} models improved as more attacks were trained, providing more data for the models to learn from.
The dynamic interaction between the attacker and defender generates diverse data for \ac{ML} training, allowing the models to adapt and handle evolving attack vectors.
Enriching the data with interaction between the defender and attacker during training resulted in better detection quality for the specific trained model.
This was especially evident in the last experiment, where data involving a defender in the attacks exhibited a more intricate relationship compared to the other two methods.
Models trained on other data were unable to comprehend the attack structure, while the model trained on defender-inclusive data performed better in handling diverse attack types.

\section{Conclusion} \label{sec:conclusion}
This study proposes a model for generating synthetic data to train \ac{ML} algorithms for intrusion detection in power grids.
The approach utilizes attack tree modeling and a game-theoretic method to create realistic data.
The attack simulation component generates attack trees while the defender minimizes the risk of attacks.
A sensitivity analysis is conducted by adjusting the number of sensors and defense funds.
The results demonstrate that \ac{RF} and \ac{XGB} achieve the best performance on the generated data.
However, the model has certain limitations, such as the requirement for knowledge of existing rules in the attack simulation component and the potential omission of certain types of attacks in the attack-defense dynamics.
Future research could explore the integration of more sophisticated \ac{ML} algorithms and more realistic attack scenarios to enhance the model's accuracy and expand its scope.
Additionally, investigating the model's effectiveness in detecting attacks on larger and more complex power grids would further improve its applicability.
Lastly, incorporating human factors, such as social engineering tactics, into the attack-defense dynamics could provide a more comprehensive understanding of cyber threats to power grids.
Nevertheless, our approach offers valuable insights to guide research in this direction.


\end{document}